\documentclass[reprint,amsmath,amssymb,aps,prl, superscriptaddress]{revtex4-1}
\usepackage[dvips]{graphicx}
\usepackage{dcolumn}
\usepackage{bm}
\usepackage{hyperref}
\usepackage{color}
\usepackage{cancel}
\usepackage{bmpsize}
\usepackage{amsmath}
\usepackage{amssymb}
\usepackage{epstopdf}
\usepackage{gensymb}
\usepackage{lipsum}
\usepackage{comment}

\usepackage{xcolor}

\newcommand{\eeqref}[1]{Eq.~(\ref{#1})}

\begin{document}

\title{Comprehensive model and performance optimization \\of phase-only spatial light modulators}

\begin{abstract}
Several spurious effects are known to degrade the performance of phase-only spatial light modulators. We introduce a comprehensive model that takes into account the major ones: curvature of the back panel, pixel crosstalk and the internal Fabry-Perot cavity. To estimate the model parameters with high accuracy, we generate blazed grating patterns and acquire the intensity response curves of the first and second diffraction orders. The quantitative model is used to generate compensating holograms, which can produce optical modes with high fidelity.

\end{abstract}

\author{A.A. Pushkina}
\affiliation{Department of Physics, University of Oxford, Oxford, OX1 3PU, UK}
\author{J.I. Costa-Filho}
\affiliation{Department of Physics, University of Oxford, Oxford, OX1 3PU, UK}
\author{G. Maltese}
\affiliation{Department of Physics, University of Oxford, Oxford, OX1 3PU, UK}
\author{A.I. Lvovsky}
\email{alex.lvovsky@physics.ox.ac.uk}
\affiliation{Department of Physics, University of Oxford, Oxford, OX1 3PU, UK}
\affiliation{Russian Quantum Center, 100 Novaya St., Skolkovo, Moscow, 143025, Russia}
\affiliation{P. N. Lebedev Physics Institute, Leninskiy prospect 53, Moscow, 119991, Russia}

\maketitle
\vspace{10 mm}
\paragraph{Introduction.}

The ability to tailor structured light beams with arbitrary intensity and phase spatial profiles is a cornerstone for a vast range of active fields, such as quantum information and communication \cite{sit2017high, parigi2015storage}, biomedical imaging \cite{angelo2017real, mcclatchy2016wide}, optical tweezing \cite{ng2010theory, arlt2001optical}, holography \cite{ren2016super}, topological photonics \cite{larocque2018reconstructing}, and metrology \cite{hermosa2014nanostep}.
In the last decades, liquid crystal on silicon spatial light modulators (LCoS SLMs) have been established as the primary tool to generate spatially structured light beams. A LCoS SLM reshapes the wavefront of an incoming beam by controlling the effective refractive index of a liquid crystal layer pixel by pixel \cite{reicherter1999optical, yan2013multicasting, osten2005evaluation}.
Among the various types of LCoS SLMs, reflective phase-only nematic SLMs are particularly popular 
\cite{bolduc2013exact, zhu2014arbitrary}. 
By making use of high-yield CMOS backplanes (pixel pitch $\sim10 {\mu}$m, fill factor up to $98\%$), high diffraction efficiencies can be achieved, while the large electro-optic coefficients of liquid crystal materials enable significant modulation depths (up to several wavelengths) and real-time operation (millisecond response time) \cite{zhang2014fundamentals}.

An ideal phase-only SLM should produce a predictable, linear and uniform phase response to the computer generated control voltage matrix. 
However, a few imperfections are known to deteriorate the SLM performance. The three most important ones are: 
the curvature of the back panel \cite{harriman2004improving}, a low finesse internal Fabry-Perot cavity \cite{xun2004phase}, and 
pixel crosstalk \cite{kulick1995electrostatic, apter2004fringing}. If not compensated, these spurious effects introduce undesirable changes to the beam wavefront.

In previous works, pixel crosstalk \cite{gemayel2016cross, persson2012reducing}, the backpanel curvature \cite{harriman2004improving, vcivzmar2010situ, zhang2012diffraction} and the cavity effect \cite{martinez2014analysis} have been studied as individual phenomena, but their joint influence on the diffracted beam has not been investigated. Yet, since these effects act simultaneously, neglecting one of them leads to imprecise estimation of the others, hindering their correct compensation.

In this work we propose a comprehensive model for all these effects  and demonstrate an effective compensation method.
To fit the model parameters, we generate  blazed grating holograms of varying amplitude and measure the position-dependent near-field intensity of the diffracted light in the first and second orders as a function of the grating amplitude.
The second order is crucial for the accurate prediction of the model parameters, since it is more sensitive to pixel crosstalk and inner cavity effects than the first order.

To demonstrate the accuracy of our model, we implement a compensation procedure valid for holograms with blazed grating patterns \cite{bolduc2013exact, davis1999encoding}.
In contrast to previously proposed compensation methods, focused on the SLM backpanel curvature, we also correct for the SLM cavity effect. By producing a cavity and curvature compensating hologram, we generate high-order Hermite-Gaussian (HG) modes of high fidelity. As an example, we demonstrate HG$_{12,12}$ with a fidelity of $94.5\%$, which is $1.9\%$ higher than the one obtained by applying a curvature only compensating hologram and $5.2\%$ higher than the one generated by a non-corrected hologram. Further, we demonstrate crosstalk correction which increases the first order diffraction efficiency by $28\%$.

\paragraph{Imperfections of a phase-only SLM.}

\begin{figure}[ht!]
\centering
\includegraphics[width=\columnwidth]{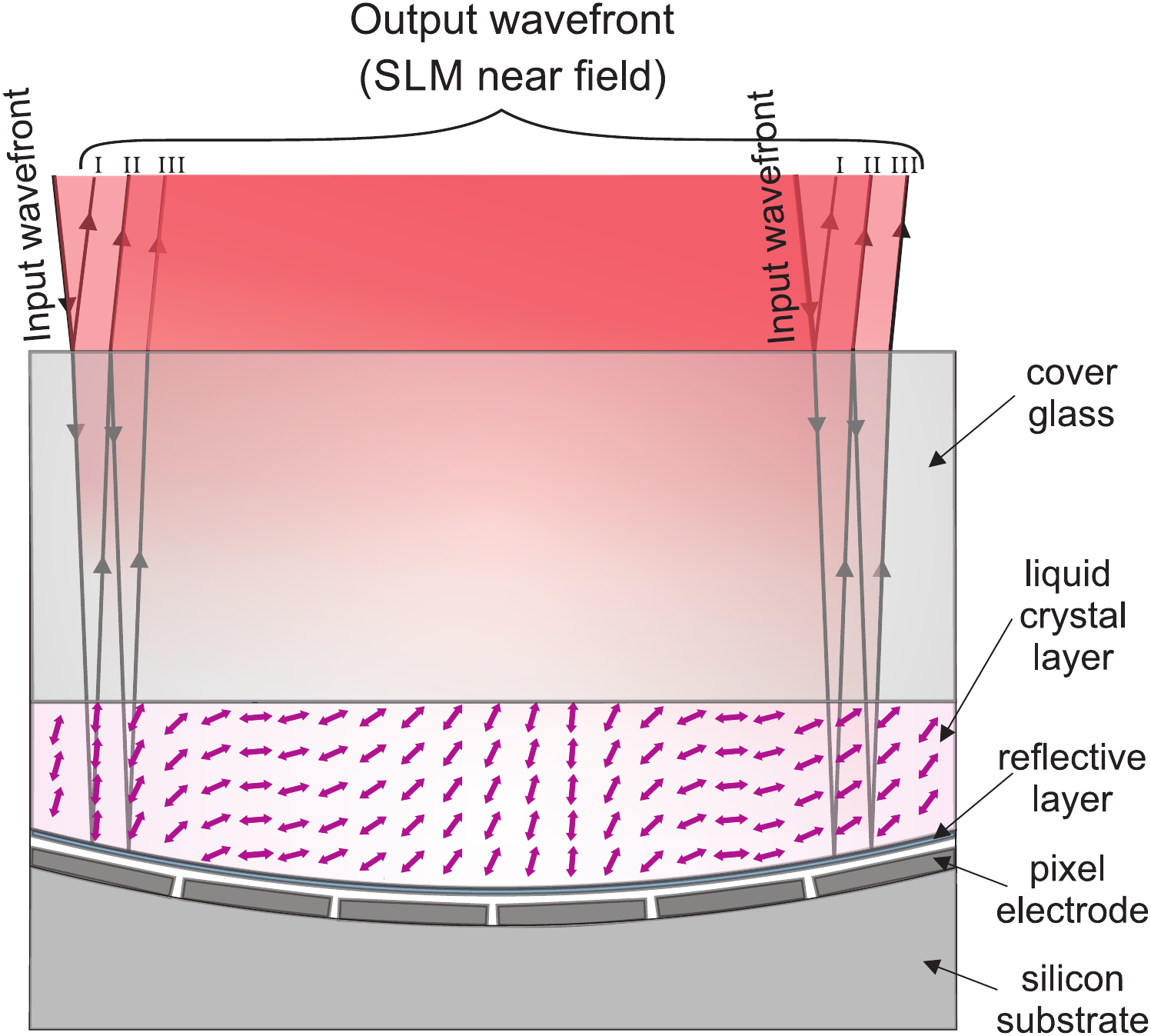}
\caption{\textbf{} Schematic cross-section of a phase-only SLM  illuminated by a wide laser beam. I, II, III outline the wavefronts produced by the Fabry-Perot cavity formed by the air-glass interface and the  backpanel.\label{SLM_scheme}}
\end{figure}

Fig.~\ref{SLM_scheme} schematizes the layered structure of a phase-only reflective SLM \cite{lazarev2019beyond}.
Due to a refractive index step at the air-glass interface, an anti-reflection coating is usually applied to the coverglass surface. The resulting coverglass reflection coefficient is low (usually around $ 0.1 $ or less \cite{martinez2014analysis}) but not negligible. 
This interface and the reflective layer of the back panel form a low-finesse Fabry-Perot cavity, which produces spurious reflection from the SLM surface.
In the following, we refer to this as the \textit{cavity effect}. We note that additional reflections may occur at the interface between the coverglass and the liquid crystal, but we found them to be negligible.

The lower part of Fig.~\ref{SLM_scheme} illustrates the curved SLM backpanel leading to a non-uniform thickness of the liquid crystal layer. Each wavefront propagating in the liquid crystal layer experiences a phase retardation given by two contributions. The first one depends on the liquid crystal molecules' orientation, determined by the pixel voltage matrix (i.e.~the printed hologram). The second contribution, voltage-independent, is determined by the additional optical path associated with the non-uniform liquid crystal thickness.

The final imperfection to be addressed is the crosstalk. Fig.~\ref{SLM_scheme} shows how it alters the orientation of the liquid crystal molecules (magenta arrows) when a 3-pixel period blazed grating is printed on the SLM. At each pixel, the liquid crystal molecules are not identically oriented, but slightly aligned with the molecules in the adjacent pixels, thereby smoothing  the phase profile experienced by an incoming wavefront.

\paragraph{Experiment.}

\begin{figure}
\includegraphics[width=\columnwidth]{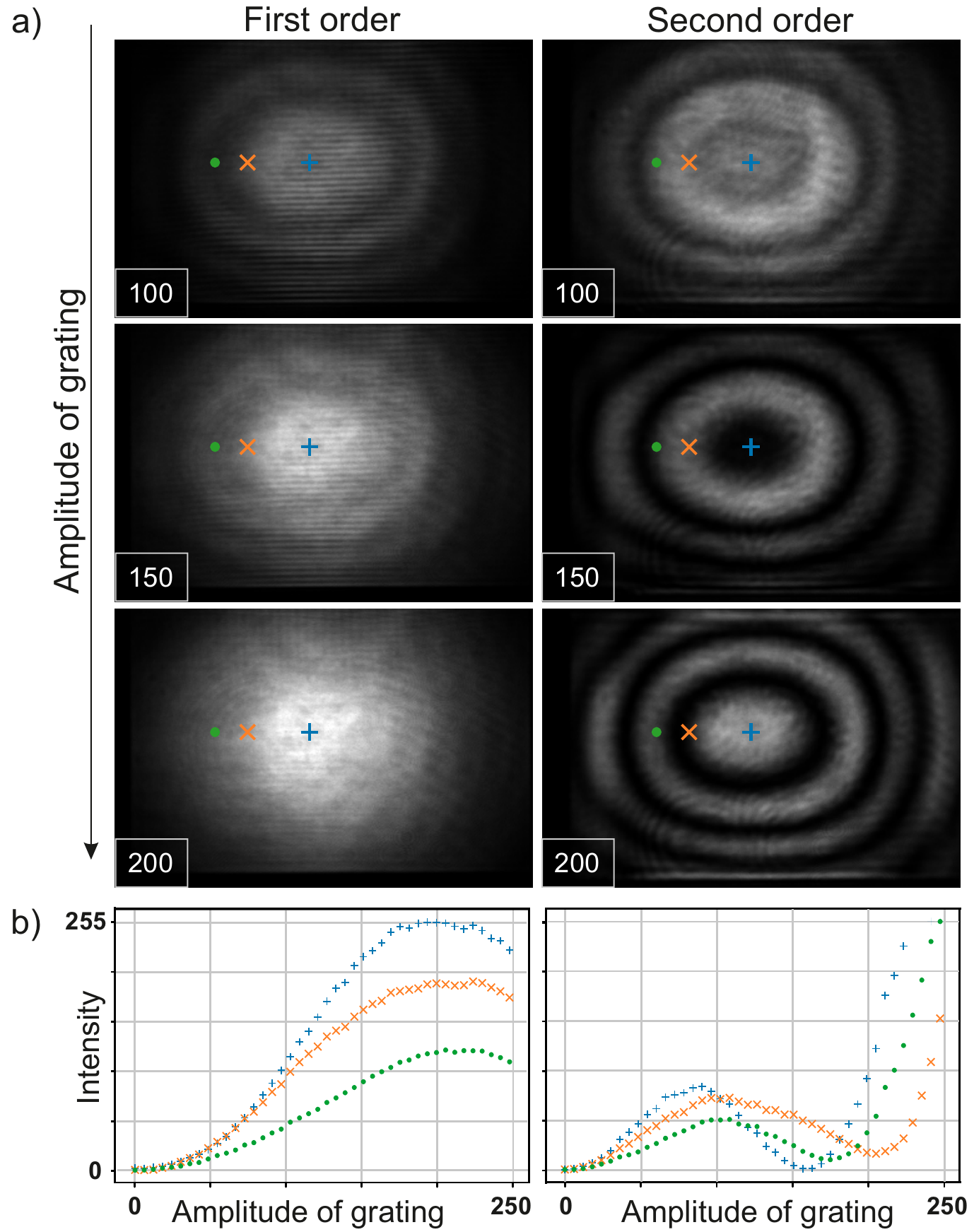}
\caption{\textbf{} (a) Near-field images of the first and second diffraction orders for three different values $a=100,150,200$ of grating amplitude. (b) Diffracted intensity versus grating amplitude for the three  SLM sections marked in Fig.~\ref{Response_curves}(a).\label{Response_curves}}
\end{figure} 

We illuminate the whole screen of a reflective phase-only LCoS-SLM (Hamamatsu X13138-02) using a continuous laser beam at 785 nm.
The SLM is placed in a parallel-aligned configuration, and the angle between the incoming and reflected beams is smaller than $5^{\circ}$. The SLM resolution is $1272\times1024$ pixels and the pixel pitch is 12.5 $\mu$m. At each pixel, the SLM is calibrated by the manufacturer to have a linear phase response to the control voltage, defined by an 8-bit integer number, commonly referred to as the \textit{gray level}. The gray level inducing a $2\pi$ phase shift is called the \textit{$2\pi$ voltage}.  

To characterize the spurious effects, we print on the SLM screen a blazed grating hologram of a 20-pixel period, varying its amplitude from 0 to 248 gray levels in steps of 2. The reflected beam is focused by a 2-inch aperture lens ($f=250$ mm) and subsequently re-imaged by another lens onto a CCD camera (UI-2140SE). An iris diaphragm in the focal plane of the first lens selects either the first or the second diffraction order. 

Fig.~\ref{Response_curves}(a) shows these images for three different values of grating amplitude. We notice an annular structure in the intensity profiles, which is a sign of the previously mentioned Fabry-Perot cavity effect. Specifically, it is a result of interference of the fields reflected from the SLM after different number of passes through the liquid crystal layer of spatially varying thickness, marked as I, II, and III in Fig.~\ref{SLM_scheme}. For high grating amplitudes, this structure is much more pronounced in the second order than in the first one. Intuitively, this is because, when the grating voltage is close to $2\pi$, the amplitude of wavefront II in the second order is greatly reduced. As a result, the amplitudes of wavefronts II and III become comparable, with their relative phase dependent on the liquid crystal layer thickness. 

This effect is further visible in Fig.~\ref{Response_curves}(b), where we plot the integrated intensity response for the three $20\times20$ sections of the SLM screen centered as marked in Fig.~\ref{Response_curves}(a). The second order intensity is low, with the behavior strongly dependent on the cavity thickness, for the grating amplitudes around the $2\pi$ voltage ($\sim200$ gray levels), resulting in well-defined rings. 



\paragraph{Theoretical model}

As illustrated in Fig.~\ref{SLM_scheme}, the optical field emitted by the SLM is given by the sum over the multiple wavefronts reflected from the cavity:
\begin{align}\label{nearfield}
E(x,y) =r+t^2\sum_{l} \left [-r e^{i\varphi(x,y)} \right ]^{l} = \frac{r+e^{i\varphi(x,y)}}{1+r e^{i\varphi(x,y)}}, 
\end{align}
where $r$ and $t$ are the reflection and transmission coefficients of the air-glass interface, $\varphi(x,y)$ the phase accumulated by the field in each round-trip and $l$ the number of round-trips. We assume  the reflection coefficient of the back panel to equal 1. The phase $\varphi(x,y)$ is given by

\begin{align}\label{Phi}
\varphi(x,y) = \theta(x,y)*g(x,y) + \alpha(x,y).
\end{align}
The first term is due to the liquid crystal's response to the applied voltage, with $\theta(x,y)$ being the phase shift in the absence of crosstalk. The crosstalk is modelled by convolving $\theta(x,y)$ with a normalized Gaussian point spread function: $g(x,y)=\mathcal{N}\left(e^{\frac{-(x^2+y^2)}{2w^2}}\right)$.  The width $w$ ranges from a fraction of one pixel to several pixels, depending on the SLM model \cite{hallstig2004fringing, gemayel2016cross}, and quantifies the strength of the effect. The second, voltage independent, term, $\alpha(x,y)$, is associated with the spatially variable thickness of the Fabry-Perot cavity.

The field distribution \eqref{nearfield} depends upon the following set of parameters: $\{r, w, a_{2\pi}(x,y), \alpha(x,y)\}$. We assume that $r$ and $w$ are spatially independent, while  $\alpha(x,y)$ and the $2\pi$ voltage $a_{2\pi}(x,y)$ are functions of transverse position. The voltage applied to the SLM corresponds to a blazed grating with its lines along the $y$ axis: 
\begin{equation}
\theta(x,y) = \frac{a}{a_{2\pi}}\cdot \bmod\left(\frac{2\pi x}{\Lambda}, 2\pi\right),
\end{equation}
where $a$ and $\Lambda$ are the grating amplitude and period, respectively.

We numerically calculate the Fourier transform of \eeqref{nearfield} and obtain the theoretical response curve of the first (second) diffraction order $I^{\rm th}_{1^{\rm st} (2^{\rm nd})}(a,x,y,\{r, w, a_{2\pi}, \alpha\})$, which is a function of the grating amplitude and  model parameters. We estimate the parameters by fitting the acquired experimental curves $I^{\rm exp}(a,x,y)$ for each $20\times20$ section of the SLM screen (examples are shown in Fig.~\ref{Response_curves}(b)).
We search for the optimal set in the following intervals: $r \in [0, 0.15]$ in steps of 0.005, $w \in [0, 2]$ pixels in steps of 0.05, $\alpha \in [0, 2\pi]$ in steps of 0.1, and $a_{2\pi}  \in [190, 220]$ in steps of 1.

We fit both diffraction orders simultaneously by considering the joint response curves $I=I_{1^{\rm st}}\cup I_{2^{\rm nd}}$. To evaluate the quality of the fit, we use chi-squared
\begin{equation}
\chi^2(x,y)= \sum_{i = 0}^{n}\frac{\left(I^{\rm exp}_i - I^{\rm th}_i\right)^2}{I^{\rm th}_i},
\end{equation}
where $n=125$ is the number of voltage values for which the data were acquired.
The following optimization procedure was used: first, for each pair of $r$ and $w$ we find the optimal spatial distribution of $\alpha(x, y)$ and $a_{2\pi}(x, y)$ by minimizing $\chi^2(x, y)$ of the joint curves for each SLM section; then, we choose the pair of $r$ and $w$ which gives the lowest value of the spatially averaged chi-squared $\chi^2_{\rm avg} = \langle\chi^2(x,y)\rangle_{x,y}$.

Fitting both orders together significantly improves the accuracy of the parameter estimation. For our SLM, the results were as follows: coverglass reflection coefficient $r=0.055$, crosstalk Gaussian point spread function width $w=0.75$ pixels, and the $2\pi$ voltage map dropping from 206 in the central part of the SLM to around 197 towards the borders.

\begin{figure}
\includegraphics[width=\columnwidth]{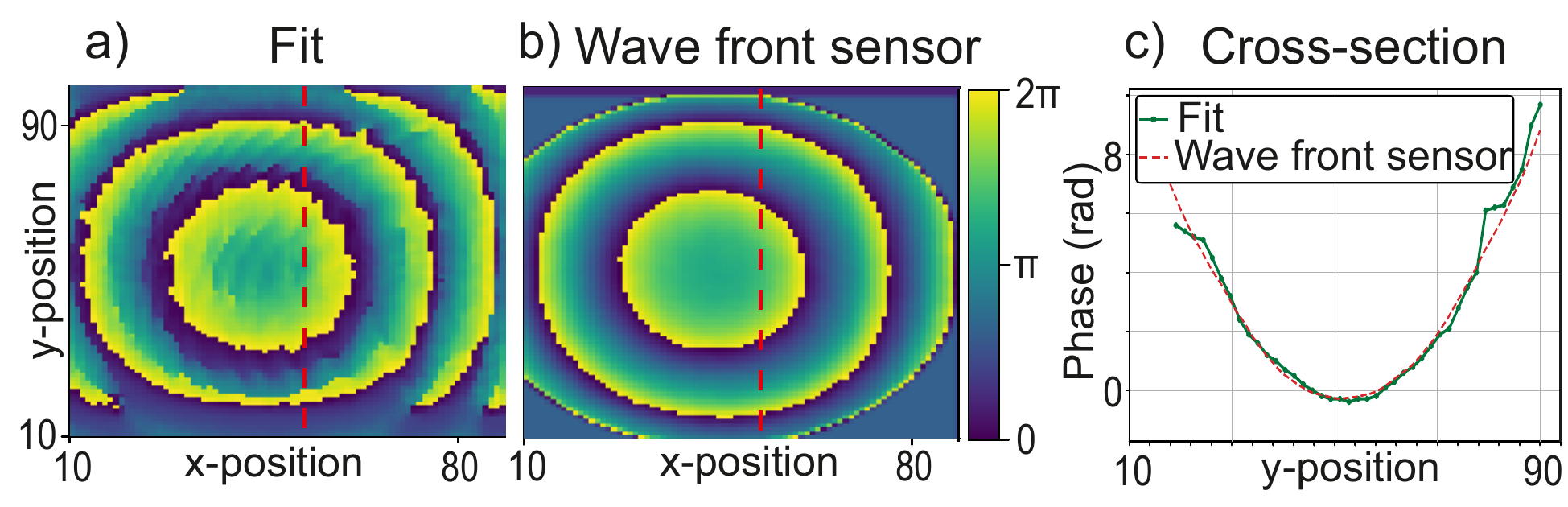}
\caption{\textbf{} Phase map $\alpha(x,y)$ (a) calculated from our model fit and (b) measured with a wavefront camera. The profiles of the cross-sections marked by dashed red lines are shown in (c).
\label{Maps}}
\end{figure}

Fig.~\ref{Maps}(a) reports $\alpha(x,y)$ resulting from the fit. As seen, $\alpha(x,y)$ resembles an elliptical paraboloid, with a peak-to-valley value of about $12$ rad ($1.9 \lambda$).
To verify the accuracy of the reconstructed $\alpha(x,y)$, we measure it using a commercial wavefront sensor camera (Phasics SID4). The fitted (Fig.~\ref{Maps}(a)) and measured (Fig.~\ref{Maps}(b)) phase maps are very similar, as evidenced further by comparing one of their cross-sections (Fig.~\ref{Maps}(c)).

\begin{figure}
\centering
\includegraphics[width=\columnwidth]{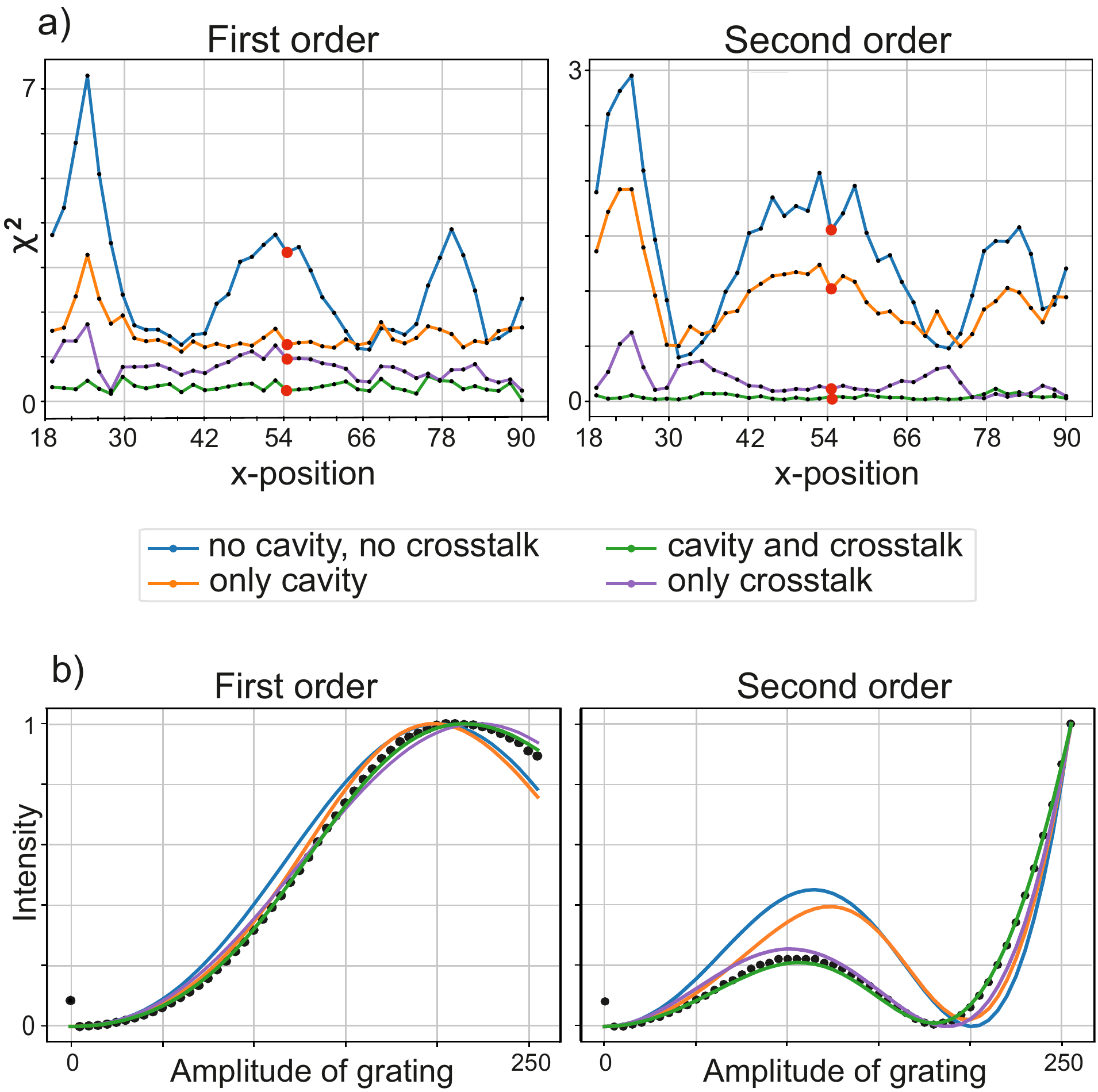}
\caption{\textbf{}
Comparison of the fit quality for the different estimation models. 
a) Fit quality parameter $\chi^2 (I^{\rm th}, I^{\rm exp})$ calculated along an SLM cross-section.
b) Measured (black dot) and fitted (coloured line) intensity response curves for an arbitrary SLM screen position (red dot in (a)).
\label{Fit_quality}}
\end{figure}

In Fig.~\ref{Fit_quality} we compare the accuracy of our model with three simpler models, in which the crosstalk and/or cavity effects are neglected, i.e. $w$ and/or $r$ are set to zero.
The model neglecting both spurious effects (blue curve) is the least accurate ($\chi^2$ is the highest). The quality of the fit improves if the model includes either cavity (orange curve) or crosstalk (purple curve), and is maximum if both effects are taken into account (green curve).  

\paragraph{Compensation of the spurious effects.}
We now address the question of how the above characterized spurious effects can be taken into account when using the SLM to produce arbitrary optical fields of amplitude $A(x,y)$ and phase $\Phi(x,y)$.
We base our approach on the widely used encoding proposed by Bolduc et al.~\cite{bolduc2013exact}, in which the pattern printed on the SLM is a modulated blazed grating of the form 
\begin{equation}\label{Psi}
\theta(x,y) = M(x,y)\cdot\bmod\left(F(x,y)+\frac{2\pi x}{\Lambda}, 2\pi\right),
\end{equation}
where $M(x,y) \in [0,1] $ and $F(x,y) \in [0, 2\pi]$ are slowly varying functions on the scale of the grating period $\Lambda$. Similarly to our experimental scheme, the field $E(x,y)$ is subjected to a direct and then inverse Fourier transform by means of two lenses. In the Fourier plane, spatial filtering is implemented to select the first diffraction order. The goal is to choose the functions $M(x,y)$ and $F(x,y)$ such that the field obtained in the image plane is the desired $A(x, y)e^{i\Phi(x, y)}$. 

For an ideal SLM modelled in Ref.~\cite{bolduc2013exact}, the transverse profile  of the field after reflection from the SLM surface is given by $E(x,y)=E_0e^{i\theta(x,y)}$, where $E_0$ is the incident field. In this case, the amplitude of $E(x,y)$ is constant, but the phase is  modulated with the period $\Lambda$, with the modulation depth and offset determined by the slowly varying $M(x,y)$ and $F(x,y)$. In an SLM with a cavity effect, however, $E(x, y)$ is given by Eq.~\eqref{nearfield}, so small amplitude modulation is also present. Importantly, the function $\alpha(x,y)$ is also slowly varying, so the reflected amplitude can still be considered quasiperiodic.

A function of this kind can be expanded into the Fourier series
\begin{equation}
E(x, y) = \sum_{k = -\infty}^{\infty}E^{(k)}_{M, F, \alpha}(x,y)e^{i2\pi kx/\Lambda},
\end{equation}
where the components
\begin{equation}\label{Fourier}
E^{(k)}_{M, F, \alpha}(x,y) = \frac{1}{\Lambda}\int_{x-\Lambda/2}^{x+\Lambda/2}E(x', y)e^{-i2\pi kx'/\Lambda}\mbox{d}x'
\end{equation}
are slowly varying functions of the transverse position. After the spatial filtering of the first diffraction order (i.e~selecting $k=1$), the field in the image plane of the SLM is given by  $E^{(1)}_{M, F, \alpha}(x,y)$. We wish this field to match the desired profile:
\begin{equation}\label{eq_for_M_and_F}
E^{(1)}_{M, F, \alpha}(x,y) = A(x, y)e^{i\Phi(x, y)}
\end{equation}
Numerically solving Eqs.~\eqref{Fourier} and \eqref{eq_for_M_and_F} to obtain $M(x,y)$ and $F(x,y)$ for the given $A(x, y)$ and $\Phi(x, y)$, as well as $\alpha(x,y)$ and $r$ known from the fit, we construct the cavity corrected hologram. Note that at this stage it is convenient to neglect the crosstalk, simplifying \eeqref{Phi} to $\varphi(x, y) = \theta(x, y) + \alpha(x, y)$.

To compensate for the crosstalk effect, we modify the hologram $\theta(x,y)$ by applying iteratively the following operation:
\begin{equation}\label{crosstalk_iter}
\theta(x,y) := 2 \theta(x,y) - \theta(x,y) * g(x,y)
\end{equation}
where $g(x,y)$ is the fitted crosstalk Gaussian kernel. 
The iterations have to be stopped when the new hologram values are about to exceed the available range of phase modulation. 
Compared to other crosstalk compensation methods, this approach is not restricted to a specific type of holograms \cite{persson2012reducing}, does not reduce the spatial resolution \cite{carbonell2019encoding} or involve complicated modelling \cite{moser2019model}. 

\begin{figure}
\includegraphics[width=0.95\columnwidth]{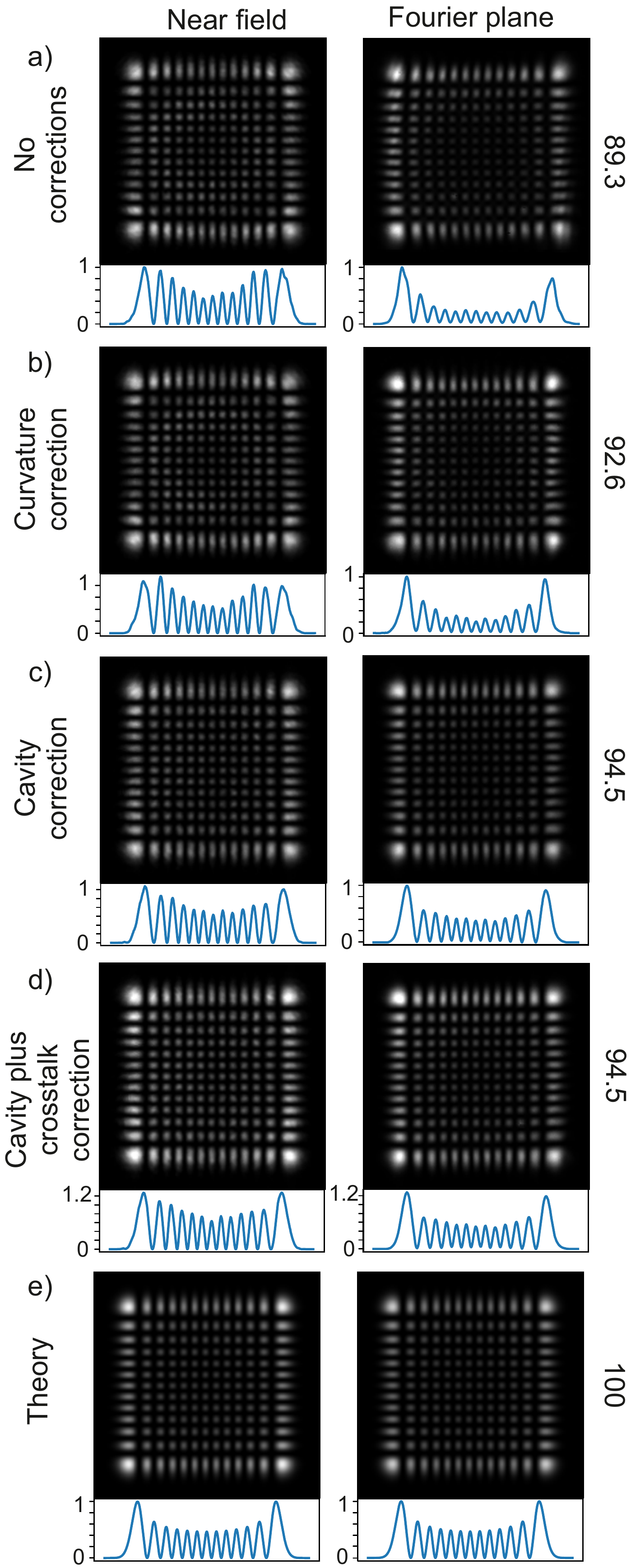}
\caption{\textbf{} Near- (left) and far-field profiles of the HG$_{12, 12}$ as produced by the SLM and measured by a camera (a-d) and expected theoretically (e). Each panel indicates the type of correction applied to the SLM hologram and the evaluated fidelity. For each image, the horizontal cross-sections through the mode centre is shown in blue line.
\label{cavity_correction}}
\end{figure}

We evaluate the proposed compensation encoding by generating  high-order Hermite-Gaussian (HG) modes. 
As a figure of merit, we evaluate the fidelity of the produced mode with respect to the ideal. For this purpose, we acquire with a camera the near and far field intensity images and apply an iterative maximum-likelihood reconstruction method \cite{lvovsky2004iterative} to obtain the first-order normalized coherence profile \begin{equation}
    g^{(1)}(x,y,x',y')=\frac{\langle E(x,y)E^*(x',y')\rangle}
    {\int\langle E(x,y)E^*(x,y)\rangle{\rm d}x{\rm d}y}.
\end{equation}
The fidelity of the experimental mode is then calculated as $$F=\int E^*_{\rm th}(x,y)E_{\rm th}(x',y')g^{(1)}(x,y,x',y'){\rm d}x{\rm d}y{\rm d}x'{\rm d}y',$$ where $E_{\rm th}(x,y)$ is the normalised theoretical profile of the ideal mode.

An example for HG$_{12,12}$ is shown in Fig.~5, comparing the theoretical mode profile with those generated by non-, partially- and fully-corrected holograms. 
The non-compensated hologram (a) produces a mode whose intensity profile has a ring-shaped modulation in the near field (especially visible in the marginal plot below the 3D photograph), due to the Fabry-Perot cavity effect, and distorted in the far field, as a consequence of the curved back panel. The fidelity of this mode with the ideal one is $89.3\%$. The curvature-corrected hologram (b) generates a mode with a higher fidelity, $92.6$\%. The far field profile is no longer distorted, but the near field intensity distribution is still modulated by the cavity interference fringes.
Hologram  (c) compensates both the curvature and cavity effects, producing a mode which best approximates the ideal mode profile (e), with a fidelity of $94.5$.

To correct for the crosstalk, we apply three iterations according to \eeqref{crosstalk_iter} to the previously calculated curvature- and cavity-compensated hologram. 
While the fidelity of the produced mode is unchanged, the diffraction efficiency increases by $28\%$, as seen in Fig~\ref{cavity_correction}(d).  The crosstalk compensation primarily modifies the areas of discontinuity in the phase profile $\theta(x,y)$ leading to a sharper blazed grating, whereas the smooth modulation functions $M(x, y)$ and $F(x, y)$ are not affected, which explains the improvement.

\paragraph{Conclusion.}
We optimize the performance of a LCoS SLM by accounting for all its major spurious effects: the curvature of the back panel, the pixel crosstalk and the low-finesse internal cavity. 
The model parameters are evaluated by measuring the intensity images of the first and second diffraction orders versus the SLM blazed grating amplitude. 
Our compensation method allows us to produce modes of significantly higher fidelity.

The presented model, characterization and compensation method can be readily applied to any phase-only LCoS SLM, with potential benefits for a vast range of applications, such as free space mode division multiplexing \cite{forbes2016creation,wang2012terabit} and maskless interference lithography \cite{behera2017design, xavier2011tunable}. 
Moreover, the ability to generate and manipulate high-order, high-fidelity Hermite-Gaussian modes may have an important impact in super-resolution imaging, paving the way towards the experimental demonstration of Hermite-Gaussian microscopy \cite{yang2016far} and related super-resolution techniques based on mode-sorting \cite{tsang2016quantum}.

\bibliography{biblio}

\begin{thebibliography}{36}%
\makeatletter
\providecommand \@ifxundefined [1]{%
 \@ifx{#1\undefined}
}%
\providecommand \@ifnum [1]{%
 \ifnum #1\expandafter \@firstoftwo
 \else \expandafter \@secondoftwo
 \fi
}%
\providecommand \@ifx [1]{%
 \ifx #1\expandafter \@firstoftwo
 \else \expandafter \@secondoftwo
 \fi
}%
\providecommand \natexlab [1]{#1}%
\providecommand \enquote  [1]{``#1''}%
\providecommand \bibnamefont  [1]{#1}%
\providecommand \bibfnamefont [1]{#1}%
\providecommand \citenamefont [1]{#1}%
\providecommand \href@noop [0]{\@secondoftwo}%
\providecommand \href [0]{\begingroup \@sanitize@url \@href}%
\providecommand \@href[1]{\@@startlink{#1}\@@href}%
\providecommand \@@href[1]{\endgroup#1\@@endlink}%
\providecommand \@sanitize@url [0]{\catcode `\\12\catcode `\$12\catcode
  `\&12\catcode `\#12\catcode `\^12\catcode `\_12\catcode `\%12\relax}%
\providecommand \@@startlink[1]{}%
\providecommand \@@endlink[0]{}%
\providecommand \url  [0]{\begingroup\@sanitize@url \@url }%
\providecommand \@url [1]{\endgroup\@href {#1}{\urlprefix }}%
\providecommand \urlprefix  [0]{URL }%
\providecommand \Eprint [0]{\href }%
\providecommand \doibase [0]{http://dx.doi.org/}%
\providecommand \selectlanguage [0]{\@gobble}%
\providecommand \bibinfo  [0]{\@secondoftwo}%
\providecommand \bibfield  [0]{\@secondoftwo}%
\providecommand \translation [1]{[#1]}%
\providecommand \BibitemOpen [0]{}%
\providecommand \bibitemStop [0]{}%
\providecommand \bibitemNoStop [0]{.\EOS\space}%
\providecommand \EOS [0]{\spacefactor3000\relax}%
\providecommand \BibitemShut  [1]{\csname bibitem#1\endcsname}%
\let\auto@bib@innerbib\@empty
\bibitem [{\citenamefont {Sit}\ \emph {et~al.}(2017)\citenamefont {Sit},
  \citenamefont {Bouchard}, \citenamefont {Fickler}, \citenamefont
  {Gagnon-Bischoff}, \citenamefont {Larocque}, \citenamefont {Heshami},
  \citenamefont {Elser}, \citenamefont {Peuntinger}, \citenamefont
  {G{\"u}nthner}, \citenamefont {Heim} \emph {et~al.}}]{sit2017high}%
  \BibitemOpen
  \bibfield  {author} {\bibinfo {author} {\bibfnamefont {A.}~\bibnamefont
  {Sit}}, \bibinfo {author} {\bibfnamefont {F.}~\bibnamefont {Bouchard}},
  \bibinfo {author} {\bibfnamefont {R.}~\bibnamefont {Fickler}}, \bibinfo
  {author} {\bibfnamefont {J.}~\bibnamefont {Gagnon-Bischoff}}, \bibinfo
  {author} {\bibfnamefont {H.}~\bibnamefont {Larocque}}, \bibinfo {author}
  {\bibfnamefont {K.}~\bibnamefont {Heshami}}, \bibinfo {author} {\bibfnamefont
  {D.}~\bibnamefont {Elser}}, \bibinfo {author} {\bibfnamefont
  {C.}~\bibnamefont {Peuntinger}}, \bibinfo {author} {\bibfnamefont
  {K.}~\bibnamefont {G{\"u}nthner}}, \bibinfo {author} {\bibfnamefont
  {B.}~\bibnamefont {Heim}},  \emph {et~al.},\ }\href@noop {} {\bibfield
  {journal} {\bibinfo  {journal} {Optica}\ }\textbf {\bibinfo {volume} {4}},\
  \bibinfo {pages} {1006} (\bibinfo {year} {2017})}\BibitemShut {NoStop}%
\bibitem [{\citenamefont {Parigi}\ \emph {et~al.}(2015)\citenamefont {Parigi},
  \citenamefont {D’Ambrosio}, \citenamefont {Arnold}, \citenamefont
  {Marrucci}, \citenamefont {Sciarrino},\ and\ \citenamefont
  {Laurat}}]{parigi2015storage}%
  \BibitemOpen
  \bibfield  {author} {\bibinfo {author} {\bibfnamefont {V.}~\bibnamefont
  {Parigi}}, \bibinfo {author} {\bibfnamefont {V.}~\bibnamefont
  {D’Ambrosio}}, \bibinfo {author} {\bibfnamefont {C.}~\bibnamefont
  {Arnold}}, \bibinfo {author} {\bibfnamefont {L.}~\bibnamefont {Marrucci}},
  \bibinfo {author} {\bibfnamefont {F.}~\bibnamefont {Sciarrino}}, \ and\
  \bibinfo {author} {\bibfnamefont {J.}~\bibnamefont {Laurat}},\ }\href@noop {}
  {\bibfield  {journal} {\bibinfo  {journal} {Nature communications}\ }\textbf
  {\bibinfo {volume} {6}},\ \bibinfo {pages} {7706} (\bibinfo {year}
  {2015})}\BibitemShut {NoStop}%
\bibitem [{\citenamefont {Angelo}\ \emph {et~al.}(2017)\citenamefont {Angelo},
  \citenamefont {Van~de Giessen},\ and\ \citenamefont
  {Gioux}}]{angelo2017real}%
  \BibitemOpen
  \bibfield  {author} {\bibinfo {author} {\bibfnamefont {J.~P.}\ \bibnamefont
  {Angelo}}, \bibinfo {author} {\bibfnamefont {M.}~\bibnamefont {Van~de
  Giessen}}, \ and\ \bibinfo {author} {\bibfnamefont {S.}~\bibnamefont
  {Gioux}},\ }\href@noop {} {\bibfield  {journal} {\bibinfo  {journal}
  {Biomedical optics express}\ }\textbf {\bibinfo {volume} {8}},\ \bibinfo
  {pages} {5113} (\bibinfo {year} {2017})}\BibitemShut {NoStop}%
\bibitem [{\citenamefont {McClatchy}\ \emph {et~al.}(2016)\citenamefont
  {McClatchy}, \citenamefont {Rizzo}, \citenamefont {Wells}, \citenamefont
  {Cheney}, \citenamefont {Hwang}, \citenamefont {Paulsen}, \citenamefont
  {Pogue},\ and\ \citenamefont {Kanick}}]{mcclatchy2016wide}%
  \BibitemOpen
  \bibfield  {author} {\bibinfo {author} {\bibfnamefont {D.~M.}\ \bibnamefont
  {McClatchy}}, \bibinfo {author} {\bibfnamefont {E.~J.}\ \bibnamefont
  {Rizzo}}, \bibinfo {author} {\bibfnamefont {W.~A.}\ \bibnamefont {Wells}},
  \bibinfo {author} {\bibfnamefont {P.~P.}\ \bibnamefont {Cheney}}, \bibinfo
  {author} {\bibfnamefont {J.~C.}\ \bibnamefont {Hwang}}, \bibinfo {author}
  {\bibfnamefont {K.~D.}\ \bibnamefont {Paulsen}}, \bibinfo {author}
  {\bibfnamefont {B.~W.}\ \bibnamefont {Pogue}}, \ and\ \bibinfo {author}
  {\bibfnamefont {S.~C.}\ \bibnamefont {Kanick}},\ }\href@noop {} {\bibfield
  {journal} {\bibinfo  {journal} {Optica}\ }\textbf {\bibinfo {volume} {3}},\
  \bibinfo {pages} {613} (\bibinfo {year} {2016})}\BibitemShut {NoStop}%
\bibitem [{\citenamefont {Ng}\ \emph {et~al.}(2010)\citenamefont {Ng},
  \citenamefont {Lin},\ and\ \citenamefont {Chan}}]{ng2010theory}%
  \BibitemOpen
  \bibfield  {author} {\bibinfo {author} {\bibfnamefont {J.}~\bibnamefont
  {Ng}}, \bibinfo {author} {\bibfnamefont {Z.}~\bibnamefont {Lin}}, \ and\
  \bibinfo {author} {\bibfnamefont {C.}~\bibnamefont {Chan}},\ }\href@noop {}
  {\bibfield  {journal} {\bibinfo  {journal} {Physical review letters}\
  }\textbf {\bibinfo {volume} {104}},\ \bibinfo {pages} {103601} (\bibinfo
  {year} {2010})}\BibitemShut {NoStop}%
\bibitem [{\citenamefont {Arlt}\ \emph {et~al.}(2001)\citenamefont {Arlt},
  \citenamefont {Garc{\'e}s-Ch{\'a}vez}, \citenamefont {Sibbett},\ and\
  \citenamefont {Dholakia}}]{arlt2001optical}%
  \BibitemOpen
  \bibfield  {author} {\bibinfo {author} {\bibfnamefont {J.}~\bibnamefont
  {Arlt}}, \bibinfo {author} {\bibfnamefont {V.}~\bibnamefont
  {Garc{\'e}s-Ch{\'a}vez}}, \bibinfo {author} {\bibfnamefont {W.}~\bibnamefont
  {Sibbett}}, \ and\ \bibinfo {author} {\bibfnamefont {K.}~\bibnamefont
  {Dholakia}},\ }\href@noop {} {\bibfield  {journal} {\bibinfo  {journal}
  {Optics Communications}\ }\textbf {\bibinfo {volume} {197}},\ \bibinfo
  {pages} {239} (\bibinfo {year} {2001})}\BibitemShut {NoStop}%
\bibitem [{\citenamefont {Ren}\ and\ \citenamefont {Lam}(2016)}]{ren2016super}%
  \BibitemOpen
  \bibfield  {author} {\bibinfo {author} {\bibfnamefont {Z.}~\bibnamefont
  {Ren}}\ and\ \bibinfo {author} {\bibfnamefont {E.~Y.}\ \bibnamefont {Lam}},\
  }in\ \href@noop {} {\emph {\bibinfo {booktitle} {Holography, Diffractive
  Optics, and Applications VII}}},\ Vol.\ \bibinfo {volume} {10022}\ (\bibinfo
  {organization} {International Society for Optics and Photonics},\ \bibinfo
  {year} {2016})\ p.\ \bibinfo {pages} {1002203}\BibitemShut {NoStop}%
\bibitem [{\citenamefont {Larocque}\ \emph {et~al.}(2018)\citenamefont
  {Larocque}, \citenamefont {Sugic}, \citenamefont {Mortimer}, \citenamefont
  {Taylor}, \citenamefont {Fickler}, \citenamefont {Boyd}, \citenamefont
  {Dennis},\ and\ \citenamefont {Karimi}}]{larocque2018reconstructing}%
  \BibitemOpen
  \bibfield  {author} {\bibinfo {author} {\bibfnamefont {H.}~\bibnamefont
  {Larocque}}, \bibinfo {author} {\bibfnamefont {D.}~\bibnamefont {Sugic}},
  \bibinfo {author} {\bibfnamefont {D.}~\bibnamefont {Mortimer}}, \bibinfo
  {author} {\bibfnamefont {A.~J.}\ \bibnamefont {Taylor}}, \bibinfo {author}
  {\bibfnamefont {R.}~\bibnamefont {Fickler}}, \bibinfo {author} {\bibfnamefont
  {R.~W.}\ \bibnamefont {Boyd}}, \bibinfo {author} {\bibfnamefont {M.~R.}\
  \bibnamefont {Dennis}}, \ and\ \bibinfo {author} {\bibfnamefont
  {E.}~\bibnamefont {Karimi}},\ }\href@noop {} {\bibfield  {journal} {\bibinfo
  {journal} {Nature Physics}\ }\textbf {\bibinfo {volume} {14}},\ \bibinfo
  {pages} {1079} (\bibinfo {year} {2018})}\BibitemShut {NoStop}%
\bibitem [{\citenamefont {Hermosa}\ \emph {et~al.}(2014)\citenamefont
  {Hermosa}, \citenamefont {Rosales-Guzm{\'a}n}, \citenamefont {Pereira},\ and\
  \citenamefont {Torres}}]{hermosa2014nanostep}%
  \BibitemOpen
  \bibfield  {author} {\bibinfo {author} {\bibfnamefont {N.}~\bibnamefont
  {Hermosa}}, \bibinfo {author} {\bibfnamefont {C.}~\bibnamefont
  {Rosales-Guzm{\'a}n}}, \bibinfo {author} {\bibfnamefont {S.}~\bibnamefont
  {Pereira}}, \ and\ \bibinfo {author} {\bibfnamefont {J.}~\bibnamefont
  {Torres}},\ }\href@noop {} {\bibfield  {journal} {\bibinfo  {journal} {Optics
  letters}\ }\textbf {\bibinfo {volume} {39}},\ \bibinfo {pages} {299}
  (\bibinfo {year} {2014})}\BibitemShut {NoStop}%
\bibitem [{\citenamefont {Reicherter}\ \emph {et~al.}(1999)\citenamefont
  {Reicherter}, \citenamefont {Haist}, \citenamefont {Wagemann},\ and\
  \citenamefont {Tiziani}}]{reicherter1999optical}%
  \BibitemOpen
  \bibfield  {author} {\bibinfo {author} {\bibfnamefont {M.}~\bibnamefont
  {Reicherter}}, \bibinfo {author} {\bibfnamefont {T.}~\bibnamefont {Haist}},
  \bibinfo {author} {\bibfnamefont {E.}~\bibnamefont {Wagemann}}, \ and\
  \bibinfo {author} {\bibfnamefont {H.~J.}\ \bibnamefont {Tiziani}},\
  }\href@noop {} {\bibfield  {journal} {\bibinfo  {journal} {Optics letters}\
  }\textbf {\bibinfo {volume} {24}},\ \bibinfo {pages} {608} (\bibinfo {year}
  {1999})}\BibitemShut {NoStop}%
\bibitem [{\citenamefont {Yan}\ \emph {et~al.}(2013)\citenamefont {Yan},
  \citenamefont {Yue}, \citenamefont {Huang}, \citenamefont {Ren},
  \citenamefont {Ahmed}, \citenamefont {Tur}, \citenamefont {Dolinar},\ and\
  \citenamefont {Willner}}]{yan2013multicasting}%
  \BibitemOpen
  \bibfield  {author} {\bibinfo {author} {\bibfnamefont {Y.}~\bibnamefont
  {Yan}}, \bibinfo {author} {\bibfnamefont {Y.}~\bibnamefont {Yue}}, \bibinfo
  {author} {\bibfnamefont {H.}~\bibnamefont {Huang}}, \bibinfo {author}
  {\bibfnamefont {Y.}~\bibnamefont {Ren}}, \bibinfo {author} {\bibfnamefont
  {N.}~\bibnamefont {Ahmed}}, \bibinfo {author} {\bibfnamefont
  {M.}~\bibnamefont {Tur}}, \bibinfo {author} {\bibfnamefont {S.}~\bibnamefont
  {Dolinar}}, \ and\ \bibinfo {author} {\bibfnamefont {A.}~\bibnamefont
  {Willner}},\ }\href@noop {} {\bibfield  {journal} {\bibinfo  {journal}
  {Optics letters}\ }\textbf {\bibinfo {volume} {38}},\ \bibinfo {pages} {3930}
  (\bibinfo {year} {2013})}\BibitemShut {NoStop}%
\bibitem [{\citenamefont {Osten}\ \emph {et~al.}(2005)\citenamefont {Osten},
  \citenamefont {Kohler}, \citenamefont {Liesener} \emph
  {et~al.}}]{osten2005evaluation}%
  \BibitemOpen
  \bibfield  {author} {\bibinfo {author} {\bibfnamefont {W.}~\bibnamefont
  {Osten}}, \bibinfo {author} {\bibfnamefont {C.}~\bibnamefont {Kohler}},
  \bibinfo {author} {\bibfnamefont {J.}~\bibnamefont {Liesener}},  \emph
  {et~al.},\ }\href@noop {} {\bibfield  {journal} {\bibinfo  {journal} {Opt.
  Pura Apl}\ }\textbf {\bibinfo {volume} {38}},\ \bibinfo {pages} {71}
  (\bibinfo {year} {2005})}\BibitemShut {NoStop}%
\bibitem [{\citenamefont {Bolduc}\ \emph {et~al.}(2013)\citenamefont {Bolduc},
  \citenamefont {Bent}, \citenamefont {Santamato}, \citenamefont {Karimi},\
  and\ \citenamefont {Boyd}}]{bolduc2013exact}%
  \BibitemOpen
  \bibfield  {author} {\bibinfo {author} {\bibfnamefont {E.}~\bibnamefont
  {Bolduc}}, \bibinfo {author} {\bibfnamefont {N.}~\bibnamefont {Bent}},
  \bibinfo {author} {\bibfnamefont {E.}~\bibnamefont {Santamato}}, \bibinfo
  {author} {\bibfnamefont {E.}~\bibnamefont {Karimi}}, \ and\ \bibinfo {author}
  {\bibfnamefont {R.~W.}\ \bibnamefont {Boyd}},\ }\href@noop {} {\bibfield
  {journal} {\bibinfo  {journal} {Optics letters}\ }\textbf {\bibinfo {volume}
  {38}},\ \bibinfo {pages} {3546} (\bibinfo {year} {2013})}\BibitemShut
  {NoStop}%
\bibitem [{\citenamefont {Zhu}\ and\ \citenamefont
  {Wang}(2014)}]{zhu2014arbitrary}%
  \BibitemOpen
  \bibfield  {author} {\bibinfo {author} {\bibfnamefont {L.}~\bibnamefont
  {Zhu}}\ and\ \bibinfo {author} {\bibfnamefont {J.}~\bibnamefont {Wang}},\
  }\href@noop {} {\bibfield  {journal} {\bibinfo  {journal} {Scientific
  reports}\ }\textbf {\bibinfo {volume} {4}},\ \bibinfo {pages} {7441}
  (\bibinfo {year} {2014})}\BibitemShut {NoStop}%
\bibitem [{\citenamefont {Zhang}\ \emph {et~al.}(2014)\citenamefont {Zhang},
  \citenamefont {You},\ and\ \citenamefont {Chu}}]{zhang2014fundamentals}%
  \BibitemOpen
  \bibfield  {author} {\bibinfo {author} {\bibfnamefont {Z.}~\bibnamefont
  {Zhang}}, \bibinfo {author} {\bibfnamefont {Z.}~\bibnamefont {You}}, \ and\
  \bibinfo {author} {\bibfnamefont {D.}~\bibnamefont {Chu}},\ }\href@noop {}
  {\bibfield  {journal} {\bibinfo  {journal} {Light: Science \& Applications}\
  }\textbf {\bibinfo {volume} {3}},\ \bibinfo {pages} {e213} (\bibinfo {year}
  {2014})}\BibitemShut {NoStop}%
\bibitem [{\citenamefont {Harriman}\ \emph {et~al.}(2004)\citenamefont
  {Harriman}, \citenamefont {Linnenberger},\ and\ \citenamefont
  {Serati}}]{harriman2004improving}%
  \BibitemOpen
  \bibfield  {author} {\bibinfo {author} {\bibfnamefont {J.~L.}\ \bibnamefont
  {Harriman}}, \bibinfo {author} {\bibfnamefont {A.}~\bibnamefont
  {Linnenberger}}, \ and\ \bibinfo {author} {\bibfnamefont {S.~A.}\
  \bibnamefont {Serati}},\ }in\ \href@noop {} {\emph {\bibinfo {booktitle}
  {Advanced Wavefront Control: Methods, Devices, and Applications II}}},\ Vol.\
  \bibinfo {volume} {5553}\ (\bibinfo {organization} {International Society for
  Optics and Photonics},\ \bibinfo {year} {2004})\ pp.\ \bibinfo {pages}
  {58--67}\BibitemShut {NoStop}%
\bibitem [{\citenamefont {Xun}\ and\ \citenamefont
  {Cohn}(2004)}]{xun2004phase}%
  \BibitemOpen
  \bibfield  {author} {\bibinfo {author} {\bibfnamefont {X.}~\bibnamefont
  {Xun}}\ and\ \bibinfo {author} {\bibfnamefont {R.~W.}\ \bibnamefont {Cohn}},\
  }\href@noop {} {\bibfield  {journal} {\bibinfo  {journal} {Applied optics}\
  }\textbf {\bibinfo {volume} {43}},\ \bibinfo {pages} {6400} (\bibinfo {year}
  {2004})}\BibitemShut {NoStop}%
\bibitem [{\citenamefont {Kulick}\ \emph {et~al.}(1995)\citenamefont {Kulick},
  \citenamefont {Jarem}, \citenamefont {Lindquist}, \citenamefont {Kowel},
  \citenamefont {Friends},\ and\ \citenamefont
  {Leslie}}]{kulick1995electrostatic}%
  \BibitemOpen
  \bibfield  {author} {\bibinfo {author} {\bibfnamefont {J.~H.}\ \bibnamefont
  {Kulick}}, \bibinfo {author} {\bibfnamefont {J.~M.}\ \bibnamefont {Jarem}},
  \bibinfo {author} {\bibfnamefont {R.~G.}\ \bibnamefont {Lindquist}}, \bibinfo
  {author} {\bibfnamefont {S.~T.}\ \bibnamefont {Kowel}}, \bibinfo {author}
  {\bibfnamefont {M.~W.}\ \bibnamefont {Friends}}, \ and\ \bibinfo {author}
  {\bibfnamefont {T.~M.}\ \bibnamefont {Leslie}},\ }\href@noop {} {\bibfield
  {journal} {\bibinfo  {journal} {Applied optics}\ }\textbf {\bibinfo {volume}
  {34}},\ \bibinfo {pages} {1901} (\bibinfo {year} {1995})}\BibitemShut
  {NoStop}%
\bibitem [{\citenamefont {Apter}\ \emph {et~al.}(2004)\citenamefont {Apter},
  \citenamefont {Efron},\ and\ \citenamefont
  {Bahat-Treidel}}]{apter2004fringing}%
  \BibitemOpen
  \bibfield  {author} {\bibinfo {author} {\bibfnamefont {B.}~\bibnamefont
  {Apter}}, \bibinfo {author} {\bibfnamefont {U.}~\bibnamefont {Efron}}, \ and\
  \bibinfo {author} {\bibfnamefont {E.}~\bibnamefont {Bahat-Treidel}},\
  }\href@noop {} {\bibfield  {journal} {\bibinfo  {journal} {Applied optics}\
  }\textbf {\bibinfo {volume} {43}},\ \bibinfo {pages} {11} (\bibinfo {year}
  {2004})}\BibitemShut {NoStop}%
\bibitem [{\citenamefont {Gemayel}\ \emph {et~al.}(2016)\citenamefont
  {Gemayel}, \citenamefont {Colicchio}, \citenamefont {Dieterlen},\ and\
  \citenamefont {Ambs}}]{gemayel2016cross}%
  \BibitemOpen
  \bibfield  {author} {\bibinfo {author} {\bibfnamefont {P.}~\bibnamefont
  {Gemayel}}, \bibinfo {author} {\bibfnamefont {B.}~\bibnamefont {Colicchio}},
  \bibinfo {author} {\bibfnamefont {A.}~\bibnamefont {Dieterlen}}, \ and\
  \bibinfo {author} {\bibfnamefont {P.}~\bibnamefont {Ambs}},\ }\href@noop {}
  {\bibfield  {journal} {\bibinfo  {journal} {Applied optics}\ }\textbf
  {\bibinfo {volume} {55}},\ \bibinfo {pages} {802} (\bibinfo {year}
  {2016})}\BibitemShut {NoStop}%
\bibitem [{\citenamefont {Persson}\ \emph {et~al.}(2012)\citenamefont
  {Persson}, \citenamefont {Engstr{\"o}m},\ and\ \citenamefont
  {Goks{\"o}r}}]{persson2012reducing}%
  \BibitemOpen
  \bibfield  {author} {\bibinfo {author} {\bibfnamefont {M.}~\bibnamefont
  {Persson}}, \bibinfo {author} {\bibfnamefont {D.}~\bibnamefont
  {Engstr{\"o}m}}, \ and\ \bibinfo {author} {\bibfnamefont {M.}~\bibnamefont
  {Goks{\"o}r}},\ }\href@noop {} {\bibfield  {journal} {\bibinfo  {journal}
  {Optics express}\ }\textbf {\bibinfo {volume} {20}},\ \bibinfo {pages}
  {22334} (\bibinfo {year} {2012})}\BibitemShut {NoStop}%
\bibitem [{\citenamefont {{\v{C}}i{\v{z}}m{\'a}r}\ \emph
  {et~al.}(2010)\citenamefont {{\v{C}}i{\v{z}}m{\'a}r}, \citenamefont
  {Mazilu},\ and\ \citenamefont {Dholakia}}]{vcivzmar2010situ}%
  \BibitemOpen
  \bibfield  {author} {\bibinfo {author} {\bibfnamefont {T.}~\bibnamefont
  {{\v{C}}i{\v{z}}m{\'a}r}}, \bibinfo {author} {\bibfnamefont {M.}~\bibnamefont
  {Mazilu}}, \ and\ \bibinfo {author} {\bibfnamefont {K.}~\bibnamefont
  {Dholakia}},\ }\href@noop {} {\bibfield  {journal} {\bibinfo  {journal}
  {Nature Photonics}\ }\textbf {\bibinfo {volume} {4}},\ \bibinfo {pages} {388}
  (\bibinfo {year} {2010})}\BibitemShut {NoStop}%
\bibitem [{\citenamefont {Zhang}\ \emph {et~al.}(2012)\citenamefont {Zhang},
  \citenamefont {Yang}, \citenamefont {Robertson}, \citenamefont {Redmond},
  \citenamefont {Pivnenko}, \citenamefont {Collings}, \citenamefont
  {Crossland},\ and\ \citenamefont {Chu}}]{zhang2012diffraction}%
  \BibitemOpen
  \bibfield  {author} {\bibinfo {author} {\bibfnamefont {Z.}~\bibnamefont
  {Zhang}}, \bibinfo {author} {\bibfnamefont {H.}~\bibnamefont {Yang}},
  \bibinfo {author} {\bibfnamefont {B.}~\bibnamefont {Robertson}}, \bibinfo
  {author} {\bibfnamefont {M.}~\bibnamefont {Redmond}}, \bibinfo {author}
  {\bibfnamefont {M.}~\bibnamefont {Pivnenko}}, \bibinfo {author}
  {\bibfnamefont {N.}~\bibnamefont {Collings}}, \bibinfo {author}
  {\bibfnamefont {W.~A.}\ \bibnamefont {Crossland}}, \ and\ \bibinfo {author}
  {\bibfnamefont {D.}~\bibnamefont {Chu}},\ }\href@noop {} {\bibfield
  {journal} {\bibinfo  {journal} {Applied optics}\ }\textbf {\bibinfo {volume}
  {51}},\ \bibinfo {pages} {3837} (\bibinfo {year} {2012})}\BibitemShut
  {NoStop}%
\bibitem [{\citenamefont {Mart{\'\i}nez}\ \emph {et~al.}(2014)\citenamefont
  {Mart{\'\i}nez}, \citenamefont {Moreno}, \citenamefont {del Mar
  S{\'a}nchez-L{\'o}pez}, \citenamefont {Vargas},\ and\ \citenamefont
  {Garc{\'\i}a-Mart{\'\i}nez}}]{martinez2014analysis}%
  \BibitemOpen
  \bibfield  {author} {\bibinfo {author} {\bibfnamefont {J.~L.}\ \bibnamefont
  {Mart{\'\i}nez}}, \bibinfo {author} {\bibfnamefont {I.}~\bibnamefont
  {Moreno}}, \bibinfo {author} {\bibfnamefont {M.}~\bibnamefont {del Mar
  S{\'a}nchez-L{\'o}pez}}, \bibinfo {author} {\bibfnamefont {A.}~\bibnamefont
  {Vargas}}, \ and\ \bibinfo {author} {\bibfnamefont {P.}~\bibnamefont
  {Garc{\'\i}a-Mart{\'\i}nez}},\ }\href@noop {} {\bibfield  {journal} {\bibinfo
   {journal} {Optics express}\ }\textbf {\bibinfo {volume} {22}},\ \bibinfo
  {pages} {25866} (\bibinfo {year} {2014})}\BibitemShut {NoStop}%
\bibitem [{\citenamefont {Davis}\ \emph {et~al.}(1999)\citenamefont {Davis},
  \citenamefont {Cottrell}, \citenamefont {Campos}, \citenamefont {Yzuel},\
  and\ \citenamefont {Moreno}}]{davis1999encoding}%
  \BibitemOpen
  \bibfield  {author} {\bibinfo {author} {\bibfnamefont {J.~A.}\ \bibnamefont
  {Davis}}, \bibinfo {author} {\bibfnamefont {D.~M.}\ \bibnamefont {Cottrell}},
  \bibinfo {author} {\bibfnamefont {J.}~\bibnamefont {Campos}}, \bibinfo
  {author} {\bibfnamefont {M.~J.}\ \bibnamefont {Yzuel}}, \ and\ \bibinfo
  {author} {\bibfnamefont {I.}~\bibnamefont {Moreno}},\ }\href@noop {}
  {\bibfield  {journal} {\bibinfo  {journal} {Applied optics}\ }\textbf
  {\bibinfo {volume} {38}},\ \bibinfo {pages} {5004} (\bibinfo {year}
  {1999})}\BibitemShut {NoStop}%
\bibitem [{\citenamefont {Lazarev}\ \emph {et~al.}(2019)\citenamefont
  {Lazarev}, \citenamefont {Chen}, \citenamefont {Strauss}, \citenamefont
  {Fontaine},\ and\ \citenamefont {Forbes}}]{lazarev2019beyond}%
  \BibitemOpen
  \bibfield  {author} {\bibinfo {author} {\bibfnamefont {G.}~\bibnamefont
  {Lazarev}}, \bibinfo {author} {\bibfnamefont {P.-J.}\ \bibnamefont {Chen}},
  \bibinfo {author} {\bibfnamefont {J.}~\bibnamefont {Strauss}}, \bibinfo
  {author} {\bibfnamefont {N.}~\bibnamefont {Fontaine}}, \ and\ \bibinfo
  {author} {\bibfnamefont {A.}~\bibnamefont {Forbes}},\ }\href@noop {}
  {\bibfield  {journal} {\bibinfo  {journal} {Optics express}\ }\textbf
  {\bibinfo {volume} {27}},\ \bibinfo {pages} {16206} (\bibinfo {year}
  {2019})}\BibitemShut {NoStop}%
\bibitem [{\citenamefont {H{\"a}llstig}\ \emph {et~al.}(2004)\citenamefont
  {H{\"a}llstig}, \citenamefont {Stigwall}, \citenamefont {Martin},
  \citenamefont {Sj{\"o}qvist},\ and\ \citenamefont
  {Lindgren}}]{hallstig2004fringing}%
  \BibitemOpen
  \bibfield  {author} {\bibinfo {author} {\bibfnamefont {E.}~\bibnamefont
  {H{\"a}llstig}}, \bibinfo {author} {\bibfnamefont {J.}~\bibnamefont
  {Stigwall}}, \bibinfo {author} {\bibfnamefont {T.}~\bibnamefont {Martin}},
  \bibinfo {author} {\bibfnamefont {L.}~\bibnamefont {Sj{\"o}qvist}}, \ and\
  \bibinfo {author} {\bibfnamefont {M.}~\bibnamefont {Lindgren}},\ }\href@noop
  {} {\bibfield  {journal} {\bibinfo  {journal} {Journal of modern optics}\
  }\textbf {\bibinfo {volume} {51}},\ \bibinfo {pages} {1233} (\bibinfo {year}
  {2004})}\BibitemShut {NoStop}%
\bibitem [{\citenamefont {Carbonell-Leal}\ and\ \citenamefont
  {Mendoza-Yero}(2019)}]{carbonell2019encoding}%
  \BibitemOpen
  \bibfield  {author} {\bibinfo {author} {\bibfnamefont {M.}~\bibnamefont
  {Carbonell-Leal}}\ and\ \bibinfo {author} {\bibfnamefont {O.}~\bibnamefont
  {Mendoza-Yero}},\ }\href@noop {} {\bibfield  {journal} {\bibinfo  {journal}
  {arXiv preprint arXiv:1903.06046}\ } (\bibinfo {year} {2019})}\BibitemShut
  {NoStop}%
\bibitem [{\citenamefont {Moser}\ \emph {et~al.}(2019)\citenamefont {Moser},
  \citenamefont {Ritsch-Marte},\ and\ \citenamefont
  {Thalhammer}}]{moser2019model}%
  \BibitemOpen
  \bibfield  {author} {\bibinfo {author} {\bibfnamefont {S.}~\bibnamefont
  {Moser}}, \bibinfo {author} {\bibfnamefont {M.}~\bibnamefont {Ritsch-Marte}},
  \ and\ \bibinfo {author} {\bibfnamefont {G.}~\bibnamefont {Thalhammer}},\
  }\href@noop {} {\bibfield  {journal} {\bibinfo  {journal} {Optics express}\
  }\textbf {\bibinfo {volume} {27}},\ \bibinfo {pages} {25046} (\bibinfo {year}
  {2019})}\BibitemShut {NoStop}%
\bibitem [{\citenamefont {Lvovsky}(2004)}]{lvovsky2004iterative}%
  \BibitemOpen
  \bibfield  {author} {\bibinfo {author} {\bibfnamefont {A.}~\bibnamefont
  {Lvovsky}},\ }\href@noop {} {\bibfield  {journal} {\bibinfo  {journal}
  {Journal of Optics B: Quantum and Semiclassical Optics}\ }\textbf {\bibinfo
  {volume} {6}},\ \bibinfo {pages} {S556} (\bibinfo {year} {2004})}\BibitemShut
  {NoStop}%
\bibitem [{\citenamefont {Forbes}\ \emph {et~al.}(2016)\citenamefont {Forbes},
  \citenamefont {Dudley},\ and\ \citenamefont {McLaren}}]{forbes2016creation}%
  \BibitemOpen
  \bibfield  {author} {\bibinfo {author} {\bibfnamefont {A.}~\bibnamefont
  {Forbes}}, \bibinfo {author} {\bibfnamefont {A.}~\bibnamefont {Dudley}}, \
  and\ \bibinfo {author} {\bibfnamefont {M.}~\bibnamefont {McLaren}},\
  }\href@noop {} {\bibfield  {journal} {\bibinfo  {journal} {Advances in Optics
  and Photonics}\ }\textbf {\bibinfo {volume} {8}},\ \bibinfo {pages} {200}
  (\bibinfo {year} {2016})}\BibitemShut {NoStop}%
\bibitem [{\citenamefont {Wang}\ \emph {et~al.}(2012)\citenamefont {Wang},
  \citenamefont {Yang}, \citenamefont {Fazal}, \citenamefont {Ahmed},
  \citenamefont {Yan}, \citenamefont {Huang}, \citenamefont {Ren},
  \citenamefont {Yue}, \citenamefont {Dolinar}, \citenamefont {Tur} \emph
  {et~al.}}]{wang2012terabit}%
  \BibitemOpen
  \bibfield  {author} {\bibinfo {author} {\bibfnamefont {J.}~\bibnamefont
  {Wang}}, \bibinfo {author} {\bibfnamefont {J.-Y.}\ \bibnamefont {Yang}},
  \bibinfo {author} {\bibfnamefont {I.~M.}\ \bibnamefont {Fazal}}, \bibinfo
  {author} {\bibfnamefont {N.}~\bibnamefont {Ahmed}}, \bibinfo {author}
  {\bibfnamefont {Y.}~\bibnamefont {Yan}}, \bibinfo {author} {\bibfnamefont
  {H.}~\bibnamefont {Huang}}, \bibinfo {author} {\bibfnamefont
  {Y.}~\bibnamefont {Ren}}, \bibinfo {author} {\bibfnamefont {Y.}~\bibnamefont
  {Yue}}, \bibinfo {author} {\bibfnamefont {S.}~\bibnamefont {Dolinar}},
  \bibinfo {author} {\bibfnamefont {M.}~\bibnamefont {Tur}},  \emph {et~al.},\
  }\href@noop {} {\bibfield  {journal} {\bibinfo  {journal} {Nature photonics}\
  }\textbf {\bibinfo {volume} {6}},\ \bibinfo {pages} {488} (\bibinfo {year}
  {2012})}\BibitemShut {NoStop}%
\bibitem [{\citenamefont {Behera}\ and\ \citenamefont
  {Joseph}(2017)}]{behera2017design}%
  \BibitemOpen
  \bibfield  {author} {\bibinfo {author} {\bibfnamefont {S.}~\bibnamefont
  {Behera}}\ and\ \bibinfo {author} {\bibfnamefont {J.}~\bibnamefont
  {Joseph}},\ }\href@noop {} {\bibfield  {journal} {\bibinfo  {journal}
  {Journal of Optics}\ }\textbf {\bibinfo {volume} {19}},\ \bibinfo {pages}
  {105103} (\bibinfo {year} {2017})}\BibitemShut {NoStop}%
\bibitem [{\citenamefont {Xavier}\ and\ \citenamefont
  {Joseph}(2011)}]{xavier2011tunable}%
  \BibitemOpen
  \bibfield  {author} {\bibinfo {author} {\bibfnamefont {J.}~\bibnamefont
  {Xavier}}\ and\ \bibinfo {author} {\bibfnamefont {J.}~\bibnamefont
  {Joseph}},\ }\href@noop {} {\bibfield  {journal} {\bibinfo  {journal} {Optics
  letters}\ }\textbf {\bibinfo {volume} {36}},\ \bibinfo {pages} {403}
  (\bibinfo {year} {2011})}\BibitemShut {NoStop}%
\bibitem [{\citenamefont {Yang}\ \emph {et~al.}(2016)\citenamefont {Yang},
  \citenamefont {Tashchilina}, \citenamefont {Moiseev}, \citenamefont {Simon},\
  and\ \citenamefont {Lvovsky}}]{yang2016far}%
  \BibitemOpen
  \bibfield  {author} {\bibinfo {author} {\bibfnamefont {F.}~\bibnamefont
  {Yang}}, \bibinfo {author} {\bibfnamefont {A.}~\bibnamefont {Tashchilina}},
  \bibinfo {author} {\bibfnamefont {E.~S.}\ \bibnamefont {Moiseev}}, \bibinfo
  {author} {\bibfnamefont {C.}~\bibnamefont {Simon}}, \ and\ \bibinfo {author}
  {\bibfnamefont {A.~I.}\ \bibnamefont {Lvovsky}},\ }\href@noop {} {\bibfield
  {journal} {\bibinfo  {journal} {Optica}\ }\textbf {\bibinfo {volume} {3}},\
  \bibinfo {pages} {1148} (\bibinfo {year} {2016})}\BibitemShut {NoStop}%
\bibitem [{\citenamefont {Tsang}\ \emph {et~al.}(2016)\citenamefont {Tsang},
  \citenamefont {Nair},\ and\ \citenamefont {Lu}}]{tsang2016quantum}%
  \BibitemOpen
  \bibfield  {author} {\bibinfo {author} {\bibfnamefont {M.}~\bibnamefont
  {Tsang}}, \bibinfo {author} {\bibfnamefont {R.}~\bibnamefont {Nair}}, \ and\
  \bibinfo {author} {\bibfnamefont {X.-M.}\ \bibnamefont {Lu}},\ }\href@noop {}
  {\bibfield  {journal} {\bibinfo  {journal} {Physical Review X}\ }\textbf
  {\bibinfo {volume} {6}},\ \bibinfo {pages} {031033} (\bibinfo {year}
  {2016})}\BibitemShut {NoStop}%
\end{thebibliography}%

\end{document}